\documentstyle[prl,preprint,aps,epsf]{revtex}
\begin{document}
\preprint{KNTP-99-03}
\bibliographystyle{plain}
\title{ Spatial Compactification and Decay-Rate Behaviour}
\author{D. K. Park$^a$, Soo-Young Lee$^a$, Hungsoo Kim$^b$}

\address{$^a$ Department of Physics, Kyungnam University,
Masan, 631-701, Korea.\\
$^b$ Department of Physics, Korea Advanced Institute of Science and
Technology, \\Taejon, 305-701, Korea.}
\date{\today}
 \maketitle

\begin{abstract}
The transition from instanton-dominated quantum tunneling regime
to sphaleron-dominated classical crossover regime is explored in 
(1+1)-dimensional scalar field theory when spatial coordinate is 
compactified. It is shown that the type of sphaleron transition
is critically dependent on the circumference of the spatial coordinate.
\end{abstract}
% \tableofcontents
% \listoffigures
\newpage
% %
Recently, much attention is paid to the winding number transition 
from instanton\cite{col79}-dominated quantum tunneling regime to the
sphaleron\cite{man83,kl84}-dominated classical crossover regime in 
$SU(2)$-Higgs model\cite{ya89,fr99-1,fr99-2,bo99}, which is believed
to describe electroweak phase transition in early universe. The active 
research in this field is mainly for the hope to understand
baryon number violating process,
which is very important consequence of electroweak chiral
anomaly\cite{ku85}.

Since, unfortunately, the sphaleron transition in $SU(2)$-Higgs model or 
real electroweak theories is too complicated to treat and it needs lot a 
numerical calculation, it is very hard to understand the real mechanism of 
the electroweak phase transition by investigating 
these models 
directly. Hence a decade ago 
Mottola and Wipf(MW)\cite{mo89} adopted a non-linear $O(3)$ model with a soft
symmetry breaking term as a toy model for the study of baryon number violating 
process. This model has an advantage that analytical expression of the
sphaleron solution can be derived by paralleling Manton's original 
argument.
Recently, the sphaleron transition in this model with and 
without Skyrme term is examined\cite{hab96,park99}.

Comparing, however, the result of Ref.\cite{park99} with that of 
Ref.\cite{fr99-1}, one can obviously conclude that MW model in itself cannot
play a key role of toy model for electroweak theory when $M_H > 6.665 M_W$,
where $M_H$ and $M_W$ are masses of Higgs and $W$ particles, respectively.
In this region $SU(2)$-Higgs theory exhibits a smooth second-order sphaleron
transition, while 
 MW model exhibits a first-order
sphaleron transition in the full range of its parameter space. Hence it may
be helpful in understanding the real nature of the electroweak phase transition
if one can find a simple toy model which exhibits both first-order and second-order
sphaleron transitions.  We argue in the 
present letter that this can be achieved by giving a nontrivial topology
to the spatial coordinate.  

If we impose a compactified spatial coordinate $x$, it naturally generates 
a periodic boundary condition $\phi(x=0) = \phi(x=L)$, 
where $\phi$ is arbitrary scalar field and $L$ is a circumference of a 
compactified spatial coordinate. On the other hand, the 
decay transition of a metastable state at finite temperature is governed
by classical configuration which satisfies another periodic boundary
condition at temporal coordinate: 
$\phi(\tau=\tau_0) = \phi(\tau=\tau_0 + 1 / T)$, where $\tau$ and $T$ are 
Euclidean time and temperature. This means we have two distinct 
periodic boundary conditions in this case, which makes the mechanism of 
the sphaleron  transition to be very complicate. 
In this letter we will explore
this issue by introducing a simple (1+1)-dimensional scalar field model and
show the type of sphaleron transition is dependent on the circumference
of the spatial coordinate.

Now, let us start with Euclidean action
\begin{equation}
\label{action}
S_E = \int d\tau dx
\left[ \frac{1}{2} \left( \frac{\partial \phi}{\partial \tau} \right)^2 +
        \frac{1}{2} \left( \frac{\partial \phi}{\partial x} \right)^2 + 
	U(\phi)  \right],
\end{equation}
where $U(\phi)$ is usual inverted double well potential
\begin{equation}
\label{potential}
U(\phi) = - \frac{\mu^2}{2 a^2} (\phi^2 - a^2)^2 + \frac{\mu^2}{2 a^2} a^4.
\end{equation}
It is very easy to show that sphaleron transition for the model 
(\ref{action}) with usual non-compactified spatial coordinate is smooth
second order if non-linear perturbation\cite{goro97} or 
number of negative modes approach\cite{lee99} are employed. 
Both approaches yield an 
identical sufficient condition for the first-order sphaleron transition and 
are very useful for the discussion of the effect of the arbitrary wall
thickness in the bubble nucleation\cite{kim99}. Fig. 1 describes  
 action-vs-temperature diagram in this simple model, which shows the type
of the sphaleron transition to be second order.

Now, let us consider action (\ref{action}) with a compactified spatial
coordinate. In this case as mentioned before 
sphaleron solution $\phi_s(x)$ must satisfy
a periodic boundary condition $\phi_s(x) = \phi_s(x+L)$.  
The explicit expression 
of $\phi_s(x)$ is\cite{man88,liang92}
\begin{equation}
\label{sphaleron}
\phi_s(x) = \frac{a}{\mu} \beta(k) \mbox{dn}[\beta(k) x, \kappa],
\end{equation}
where $k$ is modulus of elliptic function and 
\begin{eqnarray}
\label{kappa}
\kappa&=& \frac{2 \sqrt{k}}{1 + k},  \\  \nonumber
\beta(k)&=& \mu \frac{1+k}{\sqrt{1+k^2}}.
\end{eqnarray}
Since Jacobian Elliptic function $\mbox{dn}[y, \kappa]$ has period $2 K(\kappa)$, 
where $K$ is complete elliptic function\cite{byrd71}, the circumference
$L$ is defined
\begin{equation}
L_n = \frac{2 n}{\beta(k)} K(\kappa),
\end{equation}
where $n$ is some integer. Using $\phi_s(x)$ the 
classical action for sphaleron solution
is straightforwardly computed:
\begin{eqnarray}
\frac{S_n}{P}&\equiv& \int_{-L_n/2}^{L_n/2} dx
\left[ \frac{1}{2} \left(\frac{\partial \phi_s}{\partial x} \right) + U(\phi_s)
                                                                    \right]
        = n \frac{S_1}{P},  \\  \nonumber
\frac{S_1}{P}&=& \frac{a^2 \mu^2}{3 \beta(k)} \frac{(1 + k)^2}{1 + k^2}
                \left[ 4 E(\kappa) - \frac{(1-k)^2}{1+k^2} K(\kappa) \right],
\end{eqnarray}
where $P$ is period of sphaleron solution, i.e., $1/T$ and $E$ is another 
complete elliptic 
function. Since $S / P$ is interpreted 
as a barrier height of energy, the barrier height with $L = L_n$ is n-times
higher than that with $L = L_1$, and hence decay-rate is negligible for 
large $n$. In this letter, therefore, we will confine ourselves to only 
$L = L_1$ case. 

Now we apply the result of non-linear perturbation presented in 
Ref.\cite{goro97} in this model. For this we expand $\phi(x, \tau)$ around
sphaleron $\phi_s(x)$;
\begin{equation}
\phi(x, \tau) = \phi_s(x) + \eta(x, \tau),
\end{equation}
where $\eta(x, \tau)$ is small fluctuation field.
Inserting it into the equation of motion
\begin{equation}
\frac{\partial^2 \phi}{\partial \tau^2} + \frac{\partial^2 \phi}{\partial x^2}
= U^{\prime}(\phi),
\end{equation}
one can get 
\begin{equation}
\hat{l} \eta = \hat{h} \eta + G_2[\eta] + G_3[\eta],
\end{equation}
where
\begin{eqnarray}
\hat{l}&=& \frac{\partial^2}{\partial \tau^2},  \\  \nonumber
\hat{h}&=& -\frac{\partial^2}{\partial x^2} + U^{\prime \prime}(\phi_s),
                                               \\   \nonumber
G_2[\eta]&=& \frac{1}{2} U^{\prime \prime \prime}(\phi_s) \eta^2 ,
                                               \\   \nonumber
G_3[\eta]&=& \frac{1}{6} U^{\prime \prime \prime \prime}(\phi_s) \eta^3.
\end{eqnarray}
It is well-known\cite{liang92} that the eigenvalue equation of $\hat{h}$ is 
standard Lam\'{e} equation:
\begin{equation}
\frac{d^2 \psi}{dz^2} + \left[\lambda - N(N+1) \kappa^2 \mbox{sn}^2[z, \kappa]\right]
                             \psi = 0.
\end{equation}
Although the solutions of Lam\'{e} equation with period $4 K(\kappa)$ and
$2 K(\kappa)$ are well-known, solutions with other periods have not been
known yet\cite{ars64}. Since $\phi_s(x)$ has period $2 K(\kappa) / \beta(k)$,
the physically meaningful solutions of Lam\'{e} equation in this model are
those whose periods are $2 K(\kappa) / \beta(k) n$, where $n = 1, 2, 3, 
\cdots$. The eigenfunctions with period $2 K(\kappa) / \beta(k)$ and their 
corresponding eigenvalues are summarized at Table I.

The $\kappa$-dependence of eigenvalues 
$h_0$, $h_1$, and $h_2$ are shown at Fig. 2.
As shown in Fig. 2 $u_0(x)$, $u_1(x)$, and $u_2(x)$ given at Table I
are the lowest three 
eigenstates of $\hat{h}$. Although the existence of higher states is 
obvious, it is impossible to derive the eigenfunctions and their
eigenvalues analytically until now. However, the knowledge of the lowest
three eigenstates is sufficient for the discussion of the effect of 
compactified spatial coordinate in the sphaleron transition. Note that
$h_2$ is very close to zero compared to $h_0$ in the small $\kappa$ region.
We will show in the following that this effect guarantees the different types
of sphaleron transition in the small $\kappa$ region from that in large
$\kappa$ region.

The nomalization constants $C_0$, $C_1$, and $C_2$ defined at Table I
are easily derived by 
direct calculation. Since $C_1$ is not needed for further discussion, we
give only the explicit form of $C_0$ and $C_2$:
\begin{eqnarray}
C_0^2&=&\frac{3}{4}
        \frac{\beta(\kappa) \kappa^4}
	     {\frac{1}{3} K(\kappa) \left[-(2-\kappa^2) \sqrt{1-\kappa^2
	                                                     \kappa^{\prime 2}}
                                          +1-\kappa^2 + \kappa^4 \right]
	      + E(\kappa) \sqrt{1 - \kappa^2 \kappa^{\prime 2}}},
	                                          \\  \nonumber
C_2^2&=&\frac{3}{4}
        \frac{\beta(\kappa) \kappa^4}
        {\frac{1}{3} K(\kappa) \left[(2-\kappa^2) \sqrt{1-\kappa^2
                                                       \kappa^{\prime 2}}
                                            +1-\kappa^2 + \kappa^4 \right]
        - E(\kappa) \sqrt{1 - \kappa^2 \kappa^{\prime 2}}},
\end{eqnarray}
where $\kappa^{\prime 2} \equiv 1 - \kappa^2$.
Here, $\beta(\kappa)$ is $\kappa$-dependence of $\beta(k)$ whose explicit form 
can be obtained by using Eq.(\ref{kappa}):
\begin{equation}
\beta(\kappa) = \mu \sqrt{\frac{2}{2 - \kappa^2}}
               \frac{1 - \sqrt{1 - \kappa^2}}
	            {\sqrt{(2 - \kappa^2) - 2 \sqrt{1 - \kappa^2}}}.
\end{equation}
Ref.\cite{goro97} has shown that the perturbation near sphaleron solution
yields a sufficient condition for the sharp sphaleron transition.
The explicit form of this sufficient criterion is  
\begin{equation}
\label{criterion}
\frac{1}{b^2} \left[l(\omega) - l(\omega_s)\right] \equiv
<u_0 \mid f[u_0]> \hspace{.3cm} < \hspace{.3cm} 0.
\end{equation}
where $b$ is a small parameter, which is associated with small amplitude of 
periodic solution whose center is $\phi_s$ at quantum mechanical model, and
$\omega_s$ is sphaleron frequency
\begin{equation}
\omega_s \equiv \sqrt{-h_0} = \sqrt{2} \mu 
               \left[ 1 + 2 \frac{\sqrt{1 - \kappa^2 \kappa^{\prime 2}}}
	                         {2 - \kappa^2}   \right]^{1/2}.
\end{equation}
$f[u_0]$ in Eq.(\ref{criterion}) is defined as follows:
\begin{equation}
f[u_0] = - \frac{1}{2} \frac{\partial G_2}{\partial \eta} \bigg |_{\eta=u_0}
\left[ \hat{h}^{-1} + \frac{1}{2} [\hat{h} - l(2 \omega_s)]^{-1} \right]
 G_2[u_0] + \frac{3}{4} G_3[u_0]
 \end{equation}
 where $l(\omega) \equiv - \omega^2$.

 If one uses a completeness condition for $\hat{h}$, it is easy to show that 
 the condition (\ref{criterion}) reduces to 
\begin{equation}
\label{other} 
 I_0(\kappa) + I_2(\kappa) + I_{\geq 4}(\kappa) + J(\kappa) \hspace{.3cm}
                            < \hspace{.3cm} 0,
\end{equation}
where
\begin{eqnarray}
\label{def-1}
I_0(\kappa)&=& - \left[ \frac{1}{h_0} + \frac{1}{2} 
                                       \frac{1}{h_0 - l(2 \omega_s)} \right]
	      \mid <u_0 \mid G_2[u_0] > \mid^2 ,   \\  \nonumber
I_2(\kappa)&=& - \left[ \frac{1}{h_2} + \frac{1}{2} 
                                       \frac{1}{h_2 - l(2 \omega_s)} \right]
           \mid <u_2 \mid G_2[u_0] > \mid^2 ,   \\  \nonumber
I_{\geq 4}(\kappa)&=& - \sum_{n=4}^{\infty}
                \left[ \frac{1}{h_n} + \frac{1}{2} 
	                                \frac{1}{h_n - l(2 \omega_s)} \right]
              \mid <u_n \mid G_2[u_0] > \mid^2 ,   \\  \nonumber
J(\kappa)&=& \frac{3}{4} <u_0 \mid G_3[u_0]>.
\end{eqnarray}

It is worthwhile noting that the only positive one in Eq.(\ref{other})
is $I_0(\kappa)$. Now, it is clear why small $h_2$ in small $\kappa$ region
changes the type of sphaleron transition to be sharp first-order. Since
$1/h_2$ is involved in $I_2(\kappa)$ and it becomes large value in small
$\kappa$ region, the dominant contribution of the left-hand side of 
Eq. (\ref{other}) can be $I_2(\kappa)$, and hence first-order transition
may take place. 

Now let us compute $I_0(\kappa)$, $I_2(\kappa)$, and $J(\kappa)$ explicitly.
Using a recurrence relation of 
$G_n \equiv \int du \mbox{sn}^n[u, \kappa]$
\begin{equation}
G_{2m+2} = 
\frac{\mbox{sn}^{2m-1}[u, \kappa] \mbox{cn}[u, \kappa] \mbox{dn}[u, \kappa] + 2m (1 + \kappa^2)
      G_{2m} + (1 - 2m)G_{2m-2}}
      {(2m+1) \kappa^2},
\end{equation}
it is straightforward to show
\begin{equation}
\label{jcom}
J(\kappa) = -\frac{3 \mu^2 C_0^4}{a^2 \beta(\kappa)}
           \left[ A^4 K(\kappa) - 4 A^3 j_2(\kappa) + 6 A^2 j_4(\kappa)
	         -4 A j_6(\kappa) + j_8(\kappa)   \right],
\end{equation}
where
\begin{eqnarray}
A&=& \frac{1 + \kappa^2 + \sqrt{1 - \kappa^2 \kappa^{\prime 2}}}{3 \kappa^2},
                                                 \\   \nonumber
j_2(\kappa)&=& \frac{1}{\kappa^2}
              \left[K(\kappa) - E(\kappa) \right],  \\   \nonumber
j_4(\kappa)&=& \frac{1}{3 \kappa^4}
              \left[ (2 + \kappa^2) K(\kappa) - 2 (1 + \kappa^2)
	                                         E(\kappa) \right],
						 \\  \nonumber
j_6(\kappa)&=& \frac{1}{15 \kappa^6}
              \left[(8 + 3 \kappa^2 + 4 \kappa^4) K(\kappa) - 
	            (8 + 7 \kappa^2 + 8 \kappa^4) E(\kappa) \right],
		                                 \\  \nonumber
j_8(\kappa)&=& \frac{1}{105 \kappa^8}
               \left[ (48 + 16 \kappa^2 + 17 \kappa^4 + 24 \kappa^6) K(\kappa) 
	             - (48 + 40 \kappa^2 + 40 \kappa^4 + 48 \kappa^6) E(\kappa)
                                                                 \right].
\end{eqnarray}
The $\kappa$-dependence of $J(\kappa)$ is shown in Fig. 3.
Fig. 3 shows that $J(\kappa)$ in Eq. (\ref{jcom}) correctly recovers the 
$\kappa = 1$ limit of $J(\kappa)$, $-108 \mu^3 / 70 \sqrt{2} a^2$, which
can be obtained easily by calculating same quantity in the 
same model with non-compactified spatial coordinate.

Now, let us compute $I_0(\kappa)$. Using a recurrence relation of 
$D_n \equiv \int du \mbox{dn}^n[u, \kappa]$
\begin{equation}
D_{2m+3} = 
\frac{\kappa^2 \mbox{dn}^{2m}[u, \kappa] \mbox{sn}[u, \kappa] \mbox{cn}[u, \kappa] - 2 m 
      \kappa^{\prime 2} D_{2m-1} + (2m+1) (2 - \kappa^2) D_{2m+1} }
     {2(m+1)} ,
\end{equation}
and identity
$\kappa^2 \mbox{sn}^2[u, \kappa] + \mbox{dn}^2[u, \kappa] = 1$, one can show 
straightforwardly
\begin{equation}
I_0(\kappa) = - \frac{15 \pi^2 \mu^2 C_0^6}{128 a^2 h_0}
             \left[ 5 - 18 A + 24 A^2 - 16 A^3 \right]^2.
\end{equation}
Once again one can see the correct $\kappa = 1$ limit of $I_0(\kappa)$,
$3375 \pi^2 \mu^3 / 4096 \sqrt{2} a^2$ in Fig. 3.

Finally, direct computation of $I_2(\kappa)$ yields
\begin{eqnarray}
I_2(\kappa) = &-&
\frac{3 h_2 - 8 h_0}{2 h_2 (h_2 - 4 h_0)}
\frac{9 \pi^2 \mu^2}{64 a^2} C_0^4 C_2^2   \\  \nonumber
&\times&
\left[5 - 6(2A + B) + 8(A^2 + 2 A B) - 16 A^2 B \right]^2 ,
\end{eqnarray}
where
\begin{equation}
B = \frac{1 + \kappa^2 - \sqrt{1 - \kappa^2 \kappa^{\prime 2}}}{3 \kappa^2}.
\end{equation}
The fact $I_2(\kappa = 1) = 0$ as shown in Fig. 3
means that there is no correspondent discrete
mode at $\kappa = 1$. In fact, if one derives $\hat{h}$ in the same model
with non-compactified space, it is easy to show that
$\hat{h}$ becomes usual P\"{o}schl-Teller type
operator which has only two discrete modes.

Although it is impossible to calculate $I_{\geq4}(\kappa)$ analytically, we 
know that it is small negative value, which results in the following
inequality
\begin{equation}
\label{final-1}
\frac{1}{b^2} [l(\omega) - l(\omega_s)]
< I_0(\kappa) + I_2(\kappa) + J(\kappa).
\end{equation}
Fig. 4 shows $\kappa$-dependence of $\frac{a^2 \beta(\kappa)}{\mu^2 C_0^4}
\left( I_0(\kappa) + I_2(\kappa) + J(\kappa) \right)$.
Eq.(\ref{final-1}) and Fig. 4 guarantee 
the existence of $\kappa_c > \kappa^{\ast} = 
0.820621$, which distinguishes the types of transition. As expected this
simple model exhibits a first-order sphaleron transition in small
$\kappa$ region and a second-order sphaleron transition in large
$\kappa$ region which can be expected from the result of $\kappa = 1$ case.
Fig. 5 shows the numerical result of actions for the vacuum bounce and 
sphaleron solutions in the compactified model. It is worthwhile noting that
the difference between two action values becomes smaller and smaller when
the circumference of the compactified spatial coordinate decreases.
This means the possibility for the occurrence of sharp transition
is enhanced in small  $\kappa$ region, which is consistent with our 
main result.

\begin{center}
{\bf ACKNOWLEDGMENT}
\end{center}

This work was partially supported by the Kyungnam University Research Fund
in 1999.

\begin{table}

\begin{tabular}{|c|c|}  \hline 
 eigenvalues : $h_n$ & eigenfuctions : $u_n (x)$ \hspace{2.5cm} \\ \hline \hline
$h_0 = -2 \mu^2 - 4 \mu^2 \frac{\sqrt{1-\kappa^2 \kappa'^2}}{2-\kappa^2}$ &
$u_0(x) = C_0 \left[ \mbox{sn}^2[\beta (k) x, \kappa] -
 \frac{1+\kappa^2 +\sqrt{1-\kappa^2 \kappa'^2}}{ 3 \kappa^2} \right]$ 
                                   \hspace{2.0cm}  \\ \hline
$ h_1 =0$  &  $ u_1 (x) = C_1 \mbox{sn}[\beta (k) x, \kappa ] 
\mbox{cn} [ \beta (k) x, \kappa ]$ \hspace{3.2cm} \\ \hline
$h_2= -2 \mu^2 + 4 \mu^2 \frac{\sqrt{1-\kappa^2 \kappa'^2 }}{ 2 - \kappa^2}$ &
$ u_2(x) = C_2 \left[ \mbox{sn}^2[\beta (k) x, \kappa] -
 \frac{1+\kappa^2 -\sqrt{1-\kappa^2 \kappa'^2}}{ 3 \kappa^2} \right]$ 
                       \hspace{2.0cm}   \\
\end{tabular}

\vspace{0.5cm}

\caption{The lowest three eigenvalues and the corresponding 
         eigenfunctions of $\hat{h}$}

\end{table}

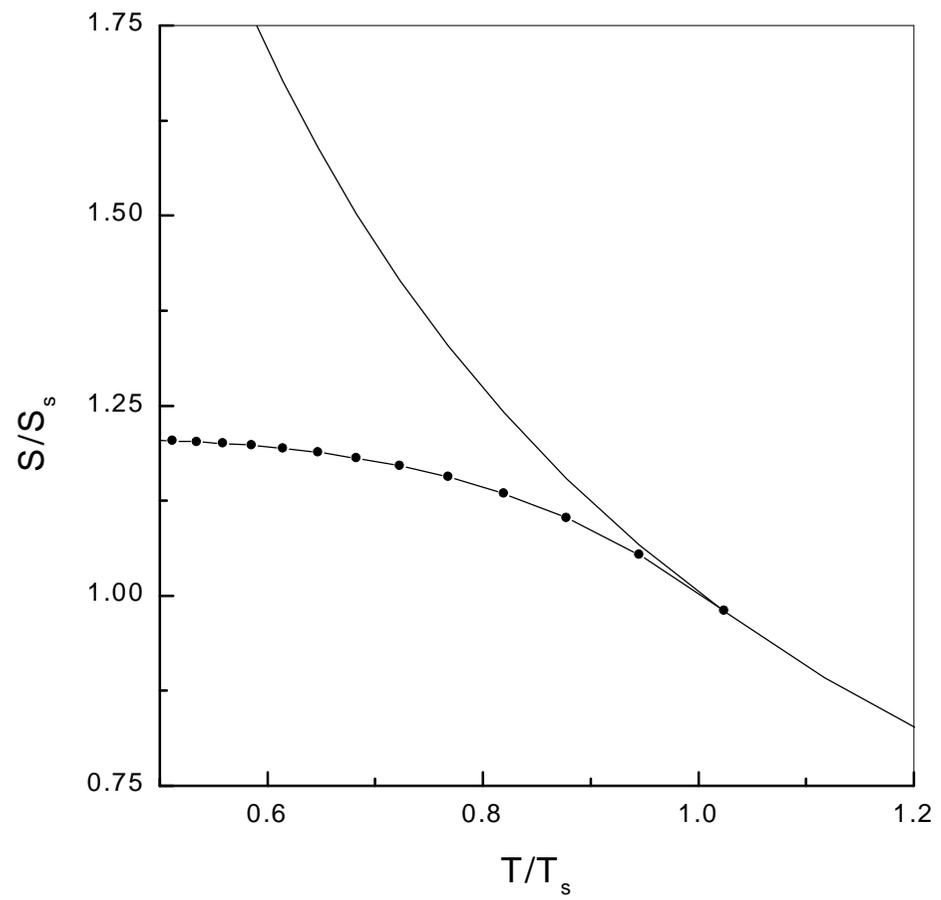
\begin{figure}
\caption{Action-vs-temperature diagram for the
action (\ref{action}) when spatial coordinate is not compactified.
 Here, $S_s$ and $1/T_s$ 
are action and period of sphaleron solution, respectively. 
The upper solid curve indicates action
for the sphaleron and the data points are for
the periodic instanton. The smooth merge at $T_s$ indicates that
the sphaleron transition is second order.
}
\end{figure}

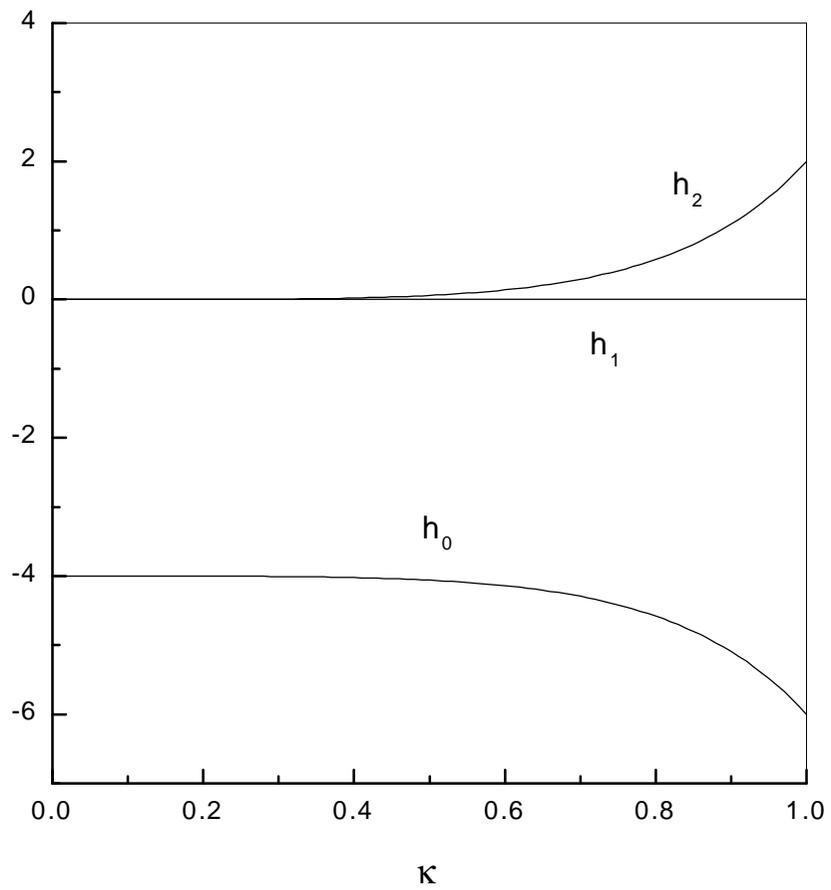
\begin{figure}
\caption{The $\kappa$-dependence of the lowest three eigenvalues of 
$\hat{h}$ when $\mu = a = 1$. The closeness of $h_2$ to zero compared to 
$h_0$ guarantees the different type of transition in the 
small $\kappa$ region from that in large $\kappa$ region.}
\end{figure}

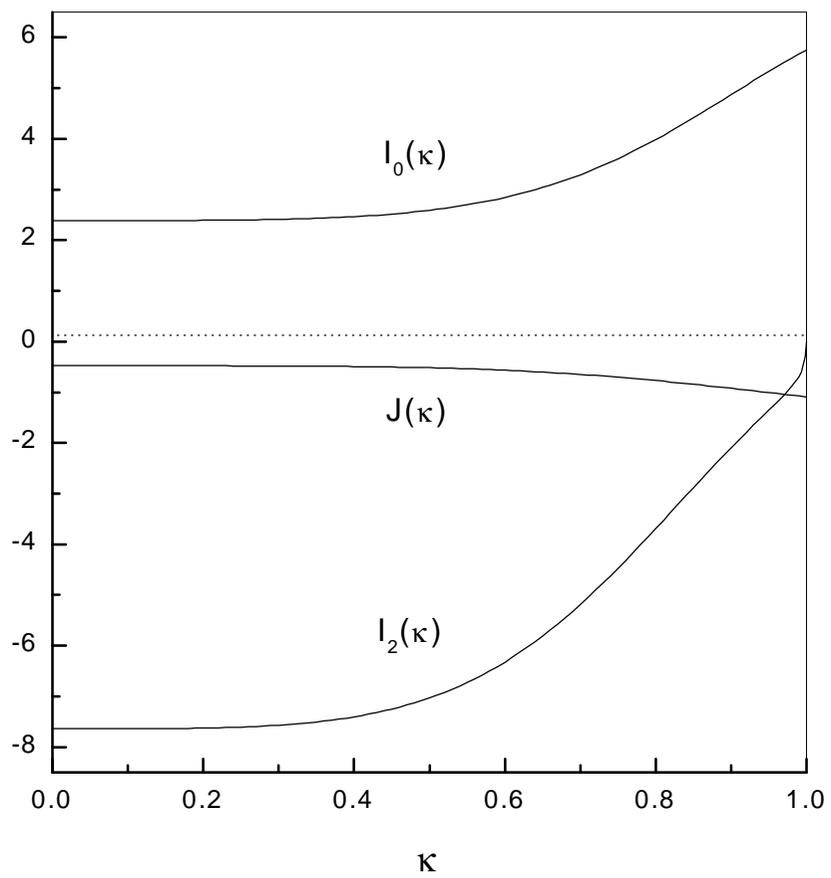
\begin{figure}
\caption{The $\kappa$-dependence of $I_0(\kappa)$, $I_2(\kappa)$, and 
$J(\kappa)$ when $\mu = a = 1$. It is easy to show that $I_0(1) = 5.74$
and $J(1) = -1.09$ are correctly recovered if one calculates the same
quantities in the non-compactified model. $I_2(1) = 0$ means that there is
no correspondent discrete mode at $\kappa = 1$ limit, which is also
easily verified by deriving $\hat{h}$ in the non-compactified model.}
\end{figure}

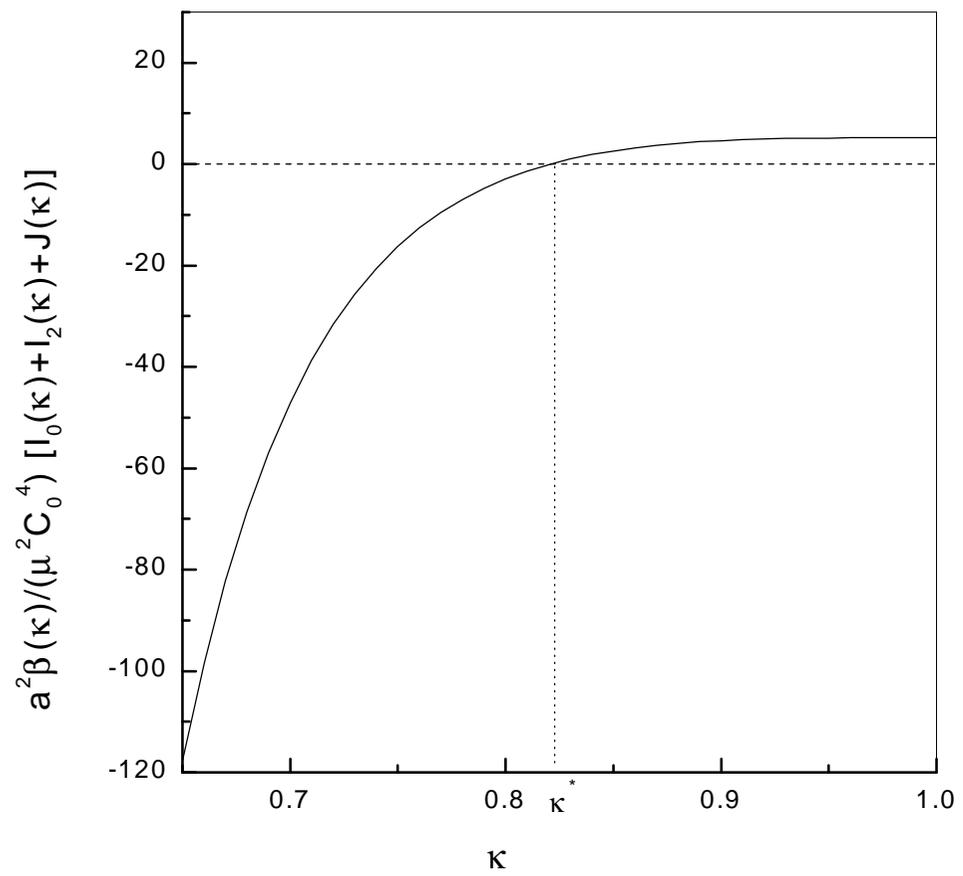
\begin{figure}
\caption{The $\kappa$-dependence of $a^2 \beta(\kappa) / \mu^2 C_0^4
\left( I_0(\kappa) + I_2(\kappa) + J(\kappa) \right)$. This figure
shows the existence of $\kappa_c > \kappa^{\ast} = 0.82061$, which 
distinguishes the types of transition.}
\end{figure}

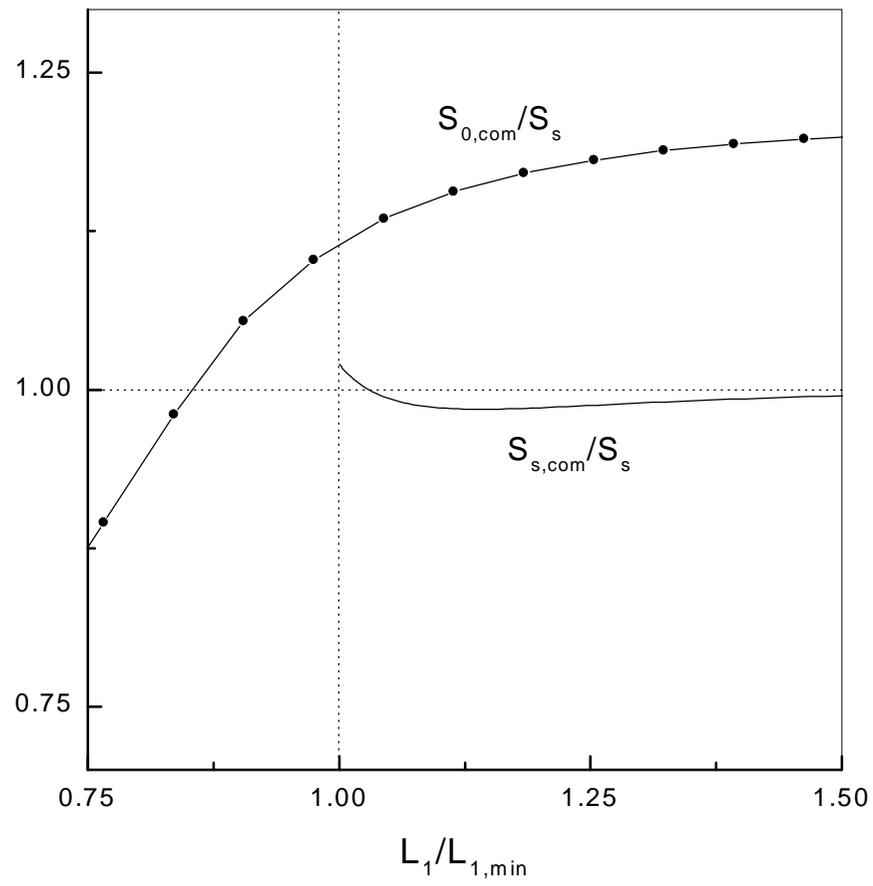
\begin{figure}
\caption{Actions for the vacuum bounce and sphaleron solutions in the 
compactified model. The fact that the difference between two action
values become smaller when the circumference of the compactified 
spatial coordinate decreases means that the possibility for the 
occurrence of sharp first-order transition increases in small
$\kappa$-region, which is consistent with our main result.}
\end{figure}

\end{document}